\documentclass[10pt]{article}

\begin{document}

\title{The equivalence principle in Kaluza-Klein gravity } 
\author{J. Ponce de Leon\thanks{E-mail: jpdel@ltp.upr.clu.edu; jpdel1@hotmail.com}\\ Laboratory of Theoretical Physics, Department of Physics\\ 
University of Puerto Rico, P.O. Box 23343, San Juan, \\ PR 00931, USA} 
\date{April 2008}

\maketitle

\begin{abstract}
In four-dimensional general relativity the spacetime outside of an isolated spherical star is described by a unique line element, which is the Schwarzschild metric. As a consequence, the ``gravitational" mass and the ``inertial" mass of a star are equal to each other. 
However,  theories that envision our world as embedded in a larger universe, with more than four dimensions, permit  a number of possible non-Schwarzschild $4D$ exteriors, which typically lead to {\it different} masses, violating the weak equivalence principle of ordinary general relativity. 
Therefore, the question arises of whether the violation of this principle, i.e., the equality of gravitational and inertial mass,  is a necessary consequence of the existence of extra dimensions. In this paper, in the context of Kaluza-Klein gravity in $5D$, we show that the answer to this question is negative.  We find a one-parameter family of asymptotically flat non-Schwarzschild static exteriors for which the inertial and  gravitational masses are equal to each other,  and equal to the Deser-Soldate mass.  This family is consistent with the Newtonian  weak-field limit as well as with the general-relativistic Schwarzschild limit. Thus, we conclude that the existence of an extra dimension, and the corresponding non-Schwarzschild exterior,  does not necessarily require different masses. However, to an observer in $4D$, it does affect the motion of test particles in $4D$, which is a consequence of the departure from the usual $(4D)$ law of geodesic motion. 

\end{abstract}

\medskip



\newpage

\section{Introduction}

According to Birkhoff's theorem, in general relativity the metric of the spacetime in the region surrounding an isolated spherical star is given by the Schwarzschild vacuum solution. The uniqueness of this solution has played an important role in our understanding of relativistic stars and gravitational collapse. Perhaps, the most significant feature of stars in general relativity is the equality between  the ``gravitational" mass $M_{g}$ and the ``inertial" mass $M_{in}$.

The ``gravitational" mass is the one  that determines the gravitational field produced by the body; this is the mass that appears in the metric tensor in a gravitational field, or, in particular,  in Newton's gravitational law. Although,  the concept of gravitational mass in a given region of space is {\it not} available in general situations, for a \underline {constant} gravitational field ($\equiv $ time-translational symmetry)  it   is possible to derive a simple expression for the total energy of matter plus field in the form of an integral over the space occupied by the matter. This is the  well-known Tolman-Whittaker formula  \cite{Landau}
\begin{equation}
\label{standard gravitational mass}
M_{g} = \int{(T^0_0 - T^1_1 - T^2_2 - T^3_3)}\sqrt{- g_{4}}dV_{3}.
\end{equation}
In the case of central symmetry of the field there is another expression for this quantity, namely 
\begin{equation}
\label{inertial mass}
m = \int{T_{0}^{0} dV_{3}}. 
\end{equation}
 Since this is the mass that acts on tests particles it  can also be obtained from the analysis of the geodesic equation.
The ``inertial" mass is the one entering in Newton's second law, it is  identified with the zeroth component of the four-momentum vector $P^{\mu}$ of the body, viz.,
\begin{equation}
\label{definition of inertial mass}
M_{in} = P^{0},
\end{equation}
and is calculated from the asymptotic behavior of the spatial part of the metric.

In fact, the equality 
\begin{equation}
\label{equal masses}
M_{g} =  M_{in} = m,
\end{equation}
constitutes a fundamental principle in general relativity and is known as the (weak) equivalence principle.  In practice, it means that objects of different nature fall at the same rate in a gravitational field.

Nowadays, there are various theories that invoke the existence of extra dimensions, although with different purposes and motivations. The study of the stellar structure, in the context of these theories,  might constitute an important approach for predicting measurable effects from the putative extra dimensions. Thus far, it has been found that the {\it effective} four-dimensional picture allows the existence of different possible, non-Schwarzschild, scenarios for the description of the spacetime outside of a spherical star. 

In braneworld models, Germani and Maartens \cite{Cristiano} have found  two exact vacuum solutions, both asymptotically Schwarzschild, which  can be used to represent the  exterior of an uniform-density star.     
Bruni, Germani and Maartens \cite{Bruni} have shown that the vacuum exterior of a collapsing dust cloud {\it cannot} be static.  Similar results have been derived by Kofinas and Papantonopoulos \cite{Kofinas} in the context of various braneworld models with curvature corrections. Other possible non-Schwarzschild stellar exteriors, with spatial spherical symmetry,  have been discussed by Dadhich {\it et al} \cite{Dadhich}, Casadio {\it et al} \cite{Casadio}, Viser and Wiltshire \cite{Viser},  and Bronnikov {\it et al} \cite{Bronnikov}. As a consequence the stellar structure in braneworld theory is rather different from that in ordinary general relativity. In particular, the gravitational and inertial masses are not necessarily equal. As an illustration, let us consider a spherical braneworld star whose region outside the surface is described by the ``spatial Schwarzschild" metric  
\begin{equation}
\label{spatialsSchw exterior}
ds^2 = \frac{1}{b^2}\left(b - 1 + \sqrt{1 - \frac{2 b {\cal{M}}}{R}}\right)^2\;dt^2 - \left( 1 - \frac{2 b {\cal{M}}}{R}\right)^{- 1}dR^2 - R^2 d\Omega^2,
\end{equation} 
which for $b = 1$ reduces to the Schwarzschild vacuum metric. The total gravitational mass of such a star is ${\cal{M}}$, i.e., $M_{g} = {\cal{M}}$. On the other hand, the total inertial mass is $M_{in} = b {\cal{M}}$; only in the Schwarzschild case $M_{g} = M_{in} = {\cal{M}}$. 

An analogous situation occurs in Kaluza-Klein gravity. Specifically, there are a number of asymptotically flat, non-Schwarzschild exteriors that can continuously be matched with the interior of a spherical star \cite{JPdeLgr-qc/0701129}. Regarding the equivalence principle, in Kaluza-Klein theory one should also consider an alternative definition of total mass, proposed by Deser and Soldate \cite{Deser},  which explicitly contains the contribution from the extra dimensions.

The object of this work is to show that the existence of an extra dimension, and non-Schwarzschild exterior, does not necessarily imply a violation of the weak equivalence principle. In order to avoid misunderstanding, we should emphasize that in the present work we examine this principle in the sense where it refers to the equality of definitions for mass (\ref{equal masses}). With this aim we test this equality within the framework of a family of static spherically symmetric metrics, which are possible static stellar exteriors \cite{JPdeLgr-qc/0701129},   in five-dimensional Kaluza-Klein gravity.  Then, we discuss and calculate the Deser-Soldate mass, say $M_{DS}$,  
 for these models. Within this framework, we find a one-parameter family of asymptotically flat non-Schwarzschild exteriors for which (as in ordinary general relativity) the masses are all equal, i.e., 
\begin{equation}
M_{g} = M_{in} = m = M_{DS}.
\end{equation}
in agreement  with the (weak) equivalence principle.

\section{Description of the model}

 In Kaluza-Klein gravity, there is a family of Ricci-flat solutions  $(R_{AB} = 0)$, which can be considered as the natural generalization of the Schwarzschild spacetime (see  a recent discussion by Lake \cite{Lake}, and references therein).  In five-dimensions, in the form given by Davidson and Owen \cite{Davidson Owen}, these solutions are described by the line element\footnote{This family of solutions has been rediscovered in different forms by Kramer \cite{Kramer} and, although in a different context,  by  Gross and Perry \cite{Gross Perry}.} 

 \begin{equation}
\label{Davidson and Owen solution}
dS^2 = \left(\frac{{\bar{a}}\rho - 1}{{\bar{a}}\rho + 1}\right)^{2\sigma k}dt^2 - \frac{1}{{\bar{a}}^4\rho^4}\frac{({\bar{a}}\rho + 1)^{2[\sigma(k - 1) + 1]}}{({\bar{a}}\rho - 1)^{2[\sigma(k - 1) - 1]}}[d\rho^2 + \rho^2d\Omega^2]
\pm  \left(\frac{{\bar{a}}\rho + 1}{{\bar{a}}\rho -1}\right)^{2\sigma}dy^2,
\end{equation}
where ${\bar{a}}$ is a constant with dimensions of $L^{- 1}$; and  $\sigma$ as well as  $k$ are parameters that obey the constraint 
\begin{equation}
\label{constraint on sigma and k}
\sigma^2(k^2 - k + 1) = 1.
\end{equation}
In the original Kaluza-Klein  theory, where the extra-dimensions are assumed to be compact and small, the higher dimensional gravity is reduced down to four dimensions by integrating out  heavier $KK$ gravitational models, see for example Ref. \cite{Segre}.  However, today it is amply accepted that the  putative extra dimension can be  large, in principle $- \infty < y < \infty$. In this new framework, the popular wisdom is that we recover our  $4D$ world  by going onto a  subspace $y$ = constant where the induced metric, say $g_{\mu\nu}^{(ind)}$, coincides with the spacetime part of (\ref{Davidson and Owen solution}). Namely, 
\begin{equation}
\label{Davidson and Owen solution on y = const.}
ds_{ind}^2 = \left(\frac{{\bar{a}}\rho - 1}{{\bar{a}}\rho + 1}\right)^{2\sigma k}dt^2 - \frac{1}{{\bar{a}}^4\rho^4}\frac{({\bar{a}}\rho + 1)^{2[\sigma(k - 1) + 1]}}{({\bar{a}}\rho - 1)^{2[\sigma(k - 1) - 1]}}[d\rho^2 + \rho^2d\Omega^2].
\end{equation}
In the limit  $k \rightarrow \infty$ $(\sigma k \rightarrow 1)$, it reproduces the Schwarzschild metric, in isotropic coordinates, for a central mass $M = 2/{\bar{a}}$, viz., 
\begin{equation}
\label{Schw. in isotropic coordinates}
ds^2_{Schw} =  ds^2_{ind|k \rightarrow \infty} = \left(\frac{1 - M/2\rho}{1 + M/2\rho}\right)^2dt^2 - \left(1 + \frac{M}{2\rho}\right)^4[d\rho^2 + \rho^2d\Omega^2].
\end{equation}
In the context of Space-Time-Matter theory (STM), metric (\ref{Davidson and Owen solution on y = const.}) is interpreted as describing extended spherical objects called solitons \cite{Wesson-JPdeL}-\cite{Liu-Overduin} (for  a recent discussion see Ref. \cite {JPdeLgr-qc/0611082} and references therein). However, in Kaluza-Klein gravity there is another possible interpretation, which is suggested by the fact that (\ref{Davidson and Owen solution on y = const.})  is singular\footnote{In general, this is a naked lightlike singularity. Only in the Schwarzschild limit  $k \rightarrow \infty$ $(\sigma k \rightarrow 1)$ the $4D$ metric (\ref{Davidson and Owen solution on y = const.}) possesses an event horizon.} at $\rho = 1/\bar{a}$, and asymptotically flat for $ \bar{a}\rho \gg 1$. Namely,  that (\ref{Davidson and Owen solution on y = const.}) can be used to describe the {\it effective} $4D$ spacetime outside of a spherical star embedded in $5D$. In this interpretation, the effective exterior is {\it not} empty because there are  nonlocal stresses induced from the Weyl curvature in $5D$, which in $4D$ behave like radiation\footnote{It should be noted that all non-Schwarzschild exteriors discussed in the literature \cite{Cristiano}-\cite{Bronnikov} have a non-vanishing effective energy-momentum tensor, which is defined through the conventional Einstein equations}.

Following this line of reasoning, the ``simplest" approach is to identify the induced metric (\ref{Davidson and Owen solution on y = const.}) with the effective $4D$ metric  outside of a spherical star, which we denote as  $g_{\mu\nu}^{(eff)}$. However, in \cite{JPdeLgr-qc/0701129} we showed that this simple approach leads to contradictions in the Newtonian limit. Therefore, we extended the discussion by considering, as a possible exterior for  a static spherical star, a family of asymptotically flat metrics which are  conformal to (\ref{Davidson and Owen solution on y = const.}).  In particular, we showed that 
\begin{equation}
\label{The exterior metric}
ds^2 = \left(\frac{\bar{a}\rho - 1}{\bar{a}\rho  + 1}\right)^{2\varepsilon}dt^2  - \frac{1}{\bar{a}^4\rho^4}\frac{(\bar{a}\rho + 1)^{2[\varepsilon + 1]}}{(\bar{a}\rho - 1)^{2[\varepsilon - 1]}}[d\rho^2 + \rho^2d\Omega^2],
\end{equation}
is the only line element, that is consistent with both: the Newtonian limit for {\it any} value of $\varepsilon$,  and the Schwarzschild limit, which is obtained for $\varepsilon = 1$. In section $3$ we will show that this line element is also consistent with the equivalence principle.

\subsection{The $5D$ solution in Schwarzschild-like coordinates}

In order to make contact with other works in the literature, it is convenient to introduce a system of coordinates where the $5D$ solutions (\ref{Davidson and Owen solution})   resemble the $4D$ Schwarzschild solution in its usual form. With this aim  we make the  coordinate transformation
\begin{equation}
r = \rho\left(1 + \frac{1}{{\bar{a}}\rho}\right)^2,
\end{equation}
which renders (\ref{Davidson and Owen solution}) into 
\begin{equation}
\label{Kramer}
dS^2 = A^a dt^2 - A^{- (a + b)}dr^2 - r^2 A^{(1 - a - b)}[d\theta^2 + \sin^2\theta d\phi^2] \pm  A^b dy^2,
\end{equation}
with
\begin{equation}
A = \left(1 - \frac{2\cal{M}}{r}\right),\;\;\;{\cal{M}} = \frac{2}{{\bar{a}}}, \;\;\;a = \sigma k,\;\;\;b = - \sigma.
\end{equation}
In this notation, the constraint (\ref{constraint on sigma and k}) reads
\begin{equation}
\label{rel. between a and b}
a^2 + ab + b^2 = 1.
\end{equation}
For the choice $a = 1$, $b = 0$ the $4D$ part of (\ref{Kramer}) becomes identical to the Schwarzschild vacuum solution for a central mass $M = {\cal{M}}$.

\subsection{The effective exterior spacetime}

For the four-dimensional interpretation of (\ref{Kramer}), as in \cite{JPdeLgr-qc/0701129} we assume that the effective metric in $4D$ is 
\begin{equation}
\label{effective 4D metric}
g_{\mu\nu}^{(eff)} = \Phi^N g_{\mu\nu}^{(ind)},
\end{equation}
where $N$ is an arbitrary real constant; $g_{\mu\nu}^{(ind)}$ is the metric induced on the subspace $y$ = constant, and $\Phi = A^{b/2}$. This {\it ansatz} is not new; it consolidates various approaches in the literature. In the context of Kaluza-Klein gravity and STM, it has been considered by Wesson \cite{Wesson book}, Kokarev \cite{Kokarev1}-\cite{Kokarev2}, Sajko \cite{Sajko}, and the present author \cite{JPdeL:gr-qc/0105120}. 
For $N = 0$, it reproduces the usual interpretation of  braneworld and STM theories. For $N = -2$ the interpretation is similar to the one provided by the canonical metric \cite{Mashhoon}. 
For $N = 1$ it yields  the classical KK interpretation \cite{Segre}, which is used, in particular,  by  Davidson-Owen \cite{Davidson Owen} and Dolan-Duff \cite{Dolan}. 

The resulting line element,  for the effective exterior $4D$ spacetime, can be written as 
\begin{equation}
\label{eff. spacetime for Kramer's metric}
ds^2_{(N)} = A^\varepsilon dt^2 - A^{[- \varepsilon + b(N -1 )]}dr^2 - r^2 A^{[(1 - \varepsilon) + b(N - 1)]}[d\theta^2 + \sin^2\theta d\phi^2],
\end{equation}
where
\begin{equation}
\label{definition of varepsilon}
\varepsilon = a + \frac{Nb}{2},
\end{equation}
and consequently, from (\ref{rel. between a and b}), we get
\begin{equation}
b_{\pm} = \frac{2\varepsilon (N - 1) \mp 2\sqrt{3(1 - \varepsilon^2) + (N - 1)^2}}{(N^2 - 2N + 4)}.
\end{equation}
We take $b = b_{+}$ for $N < 1$, and $b = b_{-}$ for $N > 1$. In this way for $\varepsilon = 1$ we recover the Schwarzschild solution. For $N = 1$, the effective $4D$ metric (\ref{eff. spacetime for Kramer's metric}) does not depend on $b$. 

For completeness, it is worth mentioning that, for a large extra dimension, both  braneworld and STM theories are mathematically equivalent \cite{STM-Brane}. In particular, they both require  slicing of the $5D$ manifold in  order to recover our $4D$ spacetime. The fundamental difference between them is that in STM there is \underline{no} matter inserted by hand into the $5D$ manifold, while in  the braneworld formalism it is postulated that there is an energy-momentum tensor in $5D$ which is discontinuous at the thin-shell (the brane) that is assumed to represent   our $4D$ universe. In what follows we continue our discussion within the context of STM.

\subsection{The stellar interior}

Standard matching conditions allow to interpret the family of asymptotically flat metrics (\ref{eff. spacetime for Kramer's metric}) as {\it possible} non-Schwarzschild scenarios for the spacetime outside of a spherical star for any $N$ and $\varepsilon $. In appendix A, we analyze the boundary conditions for the case where the stellar interior is described by a static spherical metric in curvature coordinates, viz., 

\begin{equation}
\label{int. met.}
ds^2 = e^{\omega(R)}dt^2 - e^{\sigma(R)}dR^2 - R^2 [d\theta^2 + \sin\theta d\phi^2].
\end{equation}
The boundary of the star is a ``fixed" three-dimensional surface $\Sigma$ defined by the equation $R = R_{b}$ from inside,  and $r = r_{b}$ from outside\footnote{In the notation $R_{b}$ and $r_{b}$, ``b" stands for boundary and has nothing to do with the parameter $b$ in metric (\ref{eff. spacetime for Kramer's metric}).}. The constants $R_{b}$ and $r_{b}$ are related by
\begin{equation}
R_{b} = r_{b}\left(1 - \frac{2 {\cal{M}}}{r_{b}}\right)^{\gamma},
\end{equation} 
where, in order to simplify the notation, we have set
\begin{equation}
\label{gamma}
\gamma = 1 - \varepsilon + b(N - 1).
\end{equation} 
Boundary conditions (\ref{continuity of the first fundamental form}), (\ref{cont. of sec. fund. form 1}) and (\ref{cont. of sec. fund. form 2}), for the exterior metric under consideration,  require 
\begin{equation}
\label{bond. Cond}
e^{\omega(R_{b})} = A^{\varepsilon}(r_{b}),\;\;\;e^{- \sigma(R_{b})} = A(r_{b})\left(1 + \frac{\gamma {\cal{M}}}{r_{b}A(r_{b})}\right)^2,\;\;\;\omega_{R}(R_{b}) = \frac{2\varepsilon {\cal{M}}}{r^2_{b}}\frac{A(r_{b})^{- (\gamma + 2)/2}}{[1 + \gamma {\cal{M}}/r_{b}A(r_{b})]},
\end{equation}
which impose  no restrictions on parameters $(\varepsilon, N)$.  We note that the embedding of the interior solution in $5D$ neither restricts these parameters \cite{JPdeLgr-qc/0701129}.

\subsection{Newtonian limit}

In the Newtonian limit we can write $e^{\omega (R)} = 1 + \xi f(R)$ and $e^{\sigma(R)} = 1 + \xi h(R)$, where $\xi$ is a ``small" {\it dimensionless} parameter, i.e., $|\xi| \ll  1$. To first order in $\xi$, the gravitational mass (\ref{grav. mass for the Schw. interior}) inside a sphere of radius $R$ is given by
\begin{equation}
\label{Newtonian approx for M}
M_{g}(R) = \frac{\xi R^2}{2}\left(\frac{df}{dR}\right).
\end{equation}
In this approximation, $T_{0}^{0} \gg |T_{1}^{1}|, T_{0}^{0} \gg |T_{2}^{2}| = |T_{3}^{3}|$. Then, from (\ref{standard gravitational mass}) it follows that 
$M_{g}(R) = 4\pi \int_{0}^{R}{{\bar{R}}^2\;T_{0}^{0}(\bar{R})d\bar{R}}$. Using the field equations we get
\begin{equation}
8\pi T_{0}^{0} = \frac{\xi}{R^2}\frac{d}{dR}(R\;h) + O(\xi^2).
\end{equation}
Consequently $M(R) = \xi\;R\;h/2$. On the other hand we have (\ref{Newtonian approx for M}). Thus,   in this approximation $h = R (df/dR)$. In terms of the original metric, the existence of a Newtonian limit requires.     
\begin{equation}
\label{Newtonian limit}
\label{equation for the Newtonian limit}
e^{\sigma(R)} - 1 = R \frac{d\omega}{dR} + O(\xi^2).
\end{equation}
 inside the body, including its external boundary.

\medskip

Let us now investigate in more detail the consequences of (\ref{equation for the Newtonian limit}). For this we evaluate it at the boundary and use the matching conditions (\ref{bond. Cond}). For the case under consideration, the parameter $\xi$ can be taken as $\xi = {\cal{M}}/r$. Therefore, $A = 1 - 2\xi$ and  $R = r\;[1 - 2\gamma \xi + O(\xi^2)]$. Then,  from (\ref{bond. Cond}) we  find $e^{\sigma} = 1 - 2\xi(\gamma - 1) + O(\xi^2)$ and $R\omega_{R} = 2\xi \;\varepsilon + O(\xi^2)$. Using (\ref{gamma}), we find that  the Newtonian limit (\ref{equation for the Newtonian limit}) demands 

\begin{equation}
\label{compatibility with Newtonian physics}
b(N - 1) = 0.
\end{equation}
This condition is fulfilled in two cases: (i) $b = 0$, for any arbitrary $N$ which corresponds to an exterior described by the Schwarzschild vacuum solution for $\varepsilon = 1$; (ii) $N = 1$ and $\varepsilon \neq 1$, in which case the exterior of a star is not an empty Schwarzschild spacetime.

\section{Calculating the different masses} 

In this section we evaluate  $M_{g}, M_{in}$ and $m$ mentioned in the introduction by using the standard equations in $4D$ for \underline{static} gravitational fields given by  (\ref{standard gravitational mass}) and (\ref{inertial mass}).  However, it is important to emphasize that using these equations we are \underline{not} disregarding the contribution from the ``extra" metric coefficient $g_{55} = \pm  \Phi^2 = \pm A^{b} $, because the energy-momentum tensor of the matter induced  in $4D$ is given by \cite{JPdeLgr-qc/0611082}
\begin{equation}
\label{EMT for N = 0 as an example}
8 \pi T_{\mu\nu} \sim \frac{\Phi_{\mu;\nu}}{\Phi},
\end{equation}
for any value of $N$. Thus, substituting (\ref{EMT for N = 0 as an example}) into (\ref{standard gravitational mass}) and (\ref{inertial mass}) we obtain that the effective mass measured by an observer in $4D$, is determined by the extra dimension.

\subsection{The gravitational mass}

\paragraph{Tolman-Whittaker approach:} In general relativity, when the field is {\it time-independent}, the gravitational mass inside a $3D$ volume $V_{3}$,  is given by the Tolman-Whittaker formula  (\ref{standard gravitational mass}). Using the Einstein field equations, $8\pi T_{\mu\nu} = G_{\mu\nu}$, it can be expressed in terms of the metric coefficients. 
In the present case the exterior metric has the form
\begin{equation}
\label{general exterior metric}
ds^2 = e^{\nu(r)}dt^2 - e^{\lambda(r)}dr^2 - r^2 e^{\mu(r)}[d\theta^2 + \sin^2\theta d\phi^2],
\end{equation} 
 for which we find
\begin{equation}
\label{grav. mass in terms of the metric}
M_{g}(r) = \frac{1}{2}r^2 e^{(\mu - \lambda/2 + \nu/2)}\nu',
\end{equation}
 where the prime $'$ denotes derivative with respect to $r$. We note that this is equivalent to substituting (\ref{EMT for N = 0 as an example}) into (\ref{standard gravitational mass}) \cite{JPdeLgr-qc/0611082}. Evaluating this expression for the metric (\ref{eff. spacetime for Kramer's metric})  we obtain 

\begin{equation}
\label{grav. mass}
M_{g}(r) =  \varepsilon {\cal{M}} \left(1 - \frac{2\cal{M}}{r}\right)^{b(N - 1)/2}
\end{equation}
Here, we require $\varepsilon {\cal{M}} > 0$ in order to ensure the positivity of $M$. 
Therefore, the total gravitational mass, measured by an observer located at infinity, is given by
\begin{equation}
\label{tot grav mass}
M_{g} = \varepsilon \cal{M}.
\end{equation}
\paragraph{Low velocity approximation of the geodesic equation:} This result can also be obtained from the analysis of the geodesic motion.  In fact, let us consider the radial motion of a test particle in the gravitational field described by the effective metric (\ref{eff. spacetime for Kramer's metric}). For the line element (\ref{general exterior metric}), the locally measured radial velocity $V$ and the locally measured radial acceleration $g$ are given by \cite{Bondi}
\begin{equation}
\label{local V and g}
V = \frac{e^{\lambda/2}dr}{e^{\nu/2}dt},  \;\;\;g = \frac{dV}{e^{\nu/2}dt}.
\end{equation}
From the zeroth component of the geodesic equation we get
\begin{equation}
\frac{dt}{ds} = C\;e^{-\nu},
\end{equation}
where $C$ is a constant of integration. From (\ref{general exterior metric}) evaluated at $\theta = \phi$ = constant we obtain
\begin{equation}
\label{C2}
C^{2} = \frac{e^{\nu}}{1 - V^2}.
\end{equation}
Thus, $V = \sqrt{1 - e^{\nu}/C^2}$. Taking the time derivative of this and using (\ref{C2}), we find
\begin{equation}
g = - (1 - V^2)e^{- \lambda/2}\frac{\nu'}{2},
\end{equation} 
which for the effective metric (\ref{eff. spacetime for Kramer's metric}) yields\footnote{Using (\ref{grav. mass in terms of the metric}) it takes a more familiar form. Namely,  $g = - (1 - V^2)e^{- \nu/2}\;M_{g}/R^2$.}
\begin{equation}
\label{g for moving particles}
g = - (1 - V^2)\frac{\varepsilon {\cal{M}}}{R^2}A^{[- \varepsilon + b(N- 1)]/2}.
\end{equation}
For $\varepsilon = 1$ it reproduces the usual Schwarzschild expression.  For a test body, far from the center, either at rest at a point $R$ or  in the low velocity approximation $(|V| \ll 1)$, (\ref{g for moving particles}) reduces to
\begin{equation}
g = - \frac{\varepsilon {\cal{M}}}{R^2},
\end{equation}
which is the usual Newtonian acceleration of gravity produced by a central body of gravitational mass $\varepsilon {\cal{M}}$, in agreement with (\ref{tot grav mass}).

\subsection{The inertial mass: weak field limit}

An excellent analysis of the formulae defining  the energy, momentum, and angular momentum for gravitational systems  can be found in Weinberg's book Gravitation and Cosmology \cite{Weinberg}. In appendix B, we follow that  analysis and adapt it to our\footnote{For consistency, in our research we always use the conventions and definitions found in Landau and Lifshitz \cite{Landau}. It should be noted that in the cited edition a sign has been changed in the definition of the electromagnetic field stress tensor.} notation and conventions for the signature of the metric and definition of various quantities in the theory. 

The inertial mass is the mass that appears in the four-momentum vector of the body. In particular, for an isolated  body 
\begin{equation}
P^{\mu} = \left(M_{in}, 0, 0, 0\right).
\end{equation}
So what we want  to calculate is $P^{0}$. In order to simplify the notation,  let us write the line element of the effective spacetime (\ref{eff. spacetime for Kramer's metric}) as 
\begin{equation}
\label{eff. 4D Kramer's metric with alpha and beta}
ds^2_{(N)} = A^\varepsilon dt^2 - A^{p}dr^2 - r^2 A^{q}[d\theta^2 + \sin^2\theta d\phi^2],
\end{equation}
with 
\begin{equation}
\label{original variables}
p = - \varepsilon + b(N -1 ),\;\;\;q = 1 - \varepsilon + b(N - 1),\;\;\;q - p = 1.  
\end{equation}
First we change coordinates from $(r, \theta, \phi)$ to $(x, y, z)$
\begin{equation}
x = r\sin\theta\cos\phi,\;\;\;y = r\sin\theta \sin\phi,\;\;\;z = r\cos\theta,\;\;\; r = \sqrt{x^2 + y^2 + z^2},
\end{equation} 
and denote $x = x^{1}, y = x^2, z = x^3$. Using the Minkowski metric
\begin{equation}
\label{Minkowski metric}
\eta_{\mu\nu} = diag(1, -1, -1, -1),
\end{equation} 
we have $x_{i} = - x^{i}$ and
\begin{equation}
- \eta_{ij}dx^{i}dx^{j} =   dx^2 + dy^2 + dz^2 = dr^2 + r^2[d\theta^2 + \sin^2\theta d\phi^2].
\end{equation}
In these new coordinates, and notation,  the spatial  part of the metric (\ref{eff. 4D Kramer's metric with alpha and beta}) becomes
\begin{equation}
\label{spatial metric}
g_{ij} = - A^{p}\;(1 - A)\;n_{i}n_{j} +  A^{q}\;\eta_{ij},
\end{equation}
where
\begin{equation}
n^{i} = \frac{x^i}{r},\;\;\;n_{i} = \frac{x_{i}}{r} = - \frac{\partial r}{\partial x^{i}},\;\;\;n_{i}n^{i} = - 1.
\end{equation}
In order to calculate the total mass (\ref{total mass}), we need the asymptotic behavior of the spatial part of $h_{\mu\nu}$ defined in (\ref{definition of h}). Thus,  as $r \rightarrow \infty$
\begin{equation}
\label{asympt. hij}
h_{ij} = g_{ij} - \eta_{ij}\rightarrow - \frac{2{\cal{M}}}{r}\left(n_{i}n_{j} + q\;\eta_{ij}\right) + O\;\left(\frac{1}{r^2}\right).
\end{equation} 
Also,
\begin{equation}
\label{asympt. h}
h_{i}^{i} = \eta^{jk}h_{jk} \rightarrow \frac{2{\cal{M}}(1 - 3 q)}{r} + O\;\left(\frac{1}{r^2}\right).
\end{equation}
Finally, using 
\begin{equation}
\frac{\partial n^{i}}{\partial x^{k}} = \frac{\delta^{i}_{k} + n^{i}n_{k} }{r},\;\;\;
\frac{\partial n_{i}}{\partial x^{k}} = \frac{\eta_{ik} + n_{i}n_{k} }{r},
\end{equation}
we obtain
\begin{equation}
\frac{\partial h^{j}_{k}}{\partial x^{j}} - \frac{\partial h^{j}_{j}}{\partial x^{k}} \rightarrow - \frac{4{\cal{M}}}{r^2}(1 - q)\;n_{k} + O\;\left(\frac{1}{r^3}\right).
\end{equation}
Substituting this into (\ref{total mass}) we get the total inertial mass, viz.,
\begin{equation}
M_{in} = {\cal{M}}(1 - q).
\end{equation}
Now going back to the original parameters (\ref{original variables}) we have
\begin{equation}
\label{inertial mass for Kramer's metric}
M_{in} = {\cal{M}}[\varepsilon - b(N - 1)].
\end{equation}
Thus, the inertial and gravitational masses are identical if 
\[
b(N - 1) = 0,
\]
which is the same requirement (\ref{compatibility with Newtonian physics}) needed for the compatibility of the theory with the Newtonian limit.

\subsection{The total amount of matter}

Let us now evaluate  the integral (\ref{inertial mass}) over the whole three-dimensional volume. In the present case, since the mass-energy density outside of a star does not vanish, it separates into two parts, 
\begin{equation}
\label{total m}
m  = 4\pi \int_{0}^{\infty}{R^2\;\rho dR} = 4\pi \int_{0}^{R_{b}}{\bar{R}^2\;\rho_{in}(\bar{R}) d\bar{R}} + 4\pi \int_{r_{b}}^{\infty}{R^2\;\rho_{out}(r) dR}.
\end{equation}
In the r.h.s., the first integral  represents the amount of matter\footnote{In order to avoid a misunderstanding, here by ``amount of matter" we mean the integral quantity (\ref{inertial mass}) and {\it not} the molar mass.}  inside of the star. Using that
\begin{equation}
8\pi \rho_{in} = - e^{- \sigma}\left(\frac{1}{R^2} - \frac{\sigma_{R}}{R}\right) + \frac{1}{R^2},
\end{equation}
where $\sigma_{R} = d\sigma/dR$, and $``in"$ indicates the matter quantities inside of the star, it can be written as
\begin{equation}
m_{in} = 4\pi \int_{0}^{R_{b}}{\bar{R}^2\;\rho_{in}(\bar{R}) d\bar{R}} = \frac{1}{2}\left[\bar{R}\left(1 - e^{- \sigma(\bar{R})}\right)\right]_{0}^{R_{b}} = \frac{R_{b}}{2}\left(1 - e^{- \sigma(R_{b})}\right).
\end{equation}
The second integral in the r.h.s. of (\ref{total m}) represents the amount of matter outside of the star. In order to calculate it we use $R = r e^{\mu/2}$ and 
\begin{equation}
8\pi \rho_{out} = - e^{- \lambda}\left(\mu'' + \frac{3}{4}\;\mu'^2  + \frac{3\mu'}{r} + \frac{1}{r^2} - \frac{\lambda'\mu'}{2} - \frac{\lambda'}{r}\right) + \frac{e^{- \mu}}{r^2}.
\end{equation}
We find
\begin{equation}
m_{out} = 4\pi \int_{r_{b}}^{\infty}{R^2\;\rho_{out}(r) dR} = \frac{1}{2}\left[re^{\mu/2} \left(1 - e^{\mu - \lambda}(1 + \frac{r\mu'}{2})^2\right)\right]_{r_{b}}^{\infty}.
\end{equation}
Using the boundary condition (\ref{cont. of sec. fund. form 2}) we obtain
\begin{equation}
\label{exterior mass}
m_{out} = {\cal{M}}[\varepsilon - b\;(N - 1)] - \frac{R_{b}}{2}\left[1 - e^{- \sigma(R_{b})}\right].
\end{equation}
Thus, 
\begin{equation}
\label{m, final expression}
m = m_{in} + m_{out} = {\cal{M}}\left[\varepsilon - b\;(N - 1)\right].
\end{equation}
This is identical to (\ref{inertial mass for Kramer's metric}). Thus, the total amount of matter is equivalent to the total inertial mass (\ref{inertial mass for Kramer's metric}), for any value of parameters $\varepsilon$ and $N$.   

\section{The dominant energy condition}

It should be noted that the effective matter quantities do not have to satisfy the regular energy conditions \cite{Bronnikov2}. However, in this section we point out  an interesting  link between the dominant energy condition for the effective matter and the weak equivalence principle. 

An observer in $4D$ who is not aware of the existence of an extra dimension will interpret the metric (\ref{eff. spacetime for Kramer's metric}) as if it were governed by an effective energy-momentum tensor, which is discussed in the appendix of Ref. \cite{JPdeLgr-qc/0701129}. It is easily seen that the stresses and the energy density decrease at different rates as we move outward. To simplify the discussion, we restate the asymptotic form of the effective energy-momentum tensor. As $r \rightarrow \infty$, they are   

\begin{eqnarray}
8\pi T_{0}^{0} &\rightarrow& \frac{{\cal{M}}^2\left\{(1 - \varepsilon^2) + b(N - 1)[2\varepsilon  - b(N - 1)]\right\}}{r^4}, \nonumber \\
8\pi T_{1}^{1} &\rightarrow&  - \frac{2  {\cal{M}}\; b(N - 1)}{r^3} - \frac{{\cal{M}}^2\left\{(1 - \varepsilon^2) - b(N - 1)[2 - b(N - 1) ]\right\}}{r^4}, \nonumber \\
8\pi T_{2}^{2} &\rightarrow&  \frac{ {\cal{M}}\; b (N - 1)}{r^3} + \frac{{\cal{M}}^2[(1 - \varepsilon^2) -  b(N - 1)]}{r^4}.
\end{eqnarray}
Thus, for an arbitrary $N$, the stresses decrease as $1/r^3$, whereas the energy density goes as $1/r^4$. Therefore, at large distances from the source $|T_{1}^{1}| \gg T_{0}^{0}$ and $|T_{2}^{2}| = |T_{3}^{3}| \gg T_{0}^{0}$. However, we note that if 
\begin{equation}
\label{dom. energy condition}
b(N - 1) = 0,
\end{equation}
then, the effective matter quantities satisfy the dominant energy condition ($T_{0}^{0} \geq |T_{1}^{1}|$ and $T_{0}^{0} \geq |T_{2}^{2}| = |T_{3}^{3}|$) everywhere. Thus, the equality of all masses (\ref{equal masses}) guarantees not only the existence of a Newtonian limit (\ref{compatibility with Newtonian physics}) but also the 
fulfillment of the dominant energy condition for the effective matter. 

\section{The effective exterior metric for $N = 1$ and $N = 0$}

In this section we compare and contrast the effective exteriors produced by the factorizations with $N = 1$ and $N = 0$. 

\subsection{$N = 1$}

The factorization with $N = 1$ leads to a family of non-Schwarzschild exteriors, namely
\begin{equation}
\label{exterior metric with N = 1}
ds^2_{(N = 1)} = A^\varepsilon dt^2 - A^{- \varepsilon}dr^2 - A^{(1 - \varepsilon)}r^2[d\theta^2 + \sin^2\theta d\phi^2],
\end{equation}
which are compatible with (i) the dominant energy condition; (ii) Newtonian physics, in the weak-field limit; (iii) the general-relativistic Schwarzschild limit for $\varepsilon = 1$; and (iv) the equivalence principle:
\begin{equation}
\label{equality of all masses}
M_{g} = M_{in} = m = \varepsilon {\cal{M}}. 
\end{equation}
The only theoretical restriction on $\varepsilon $ comes from the positivity of the energy-density outside of the star. Namely, 
\begin{equation}
\label{exterior matter density}
8\pi \rho_{out} = \frac{{\cal{M}}^2(1 - \varepsilon^2)}{r^4}\left(1 - \frac{2\cal{M}}{r}\right)^{(\varepsilon - 2)},\;\;\;0 \leq \varepsilon \leq 1.
\end{equation}
However, experimental and observational evidence \cite{Overduin}, \cite{Liu-Overduin} suggests that $\varepsilon$ should be very close to $1$, which is its Schwarzschild value. 

The effective exterior with $N = 1$ presents another outstanding feature. Specifically, that for the exterior matter distribution, 
\begin{equation}
\label{exotic exterior}
{(T^0_0 - T^1_1 - T^2_2 - T^3_3)} = 0,
\end{equation}
because $T^1_1  = - T^0_0 $ and  $T^1_1 = T^2_2 = T^0_0 $. Thus, although the exterior (\ref{exterior metric with N = 1}) is not empty, it shares a common property with the Schwarzschild vacuum exterior, namely, that it does not contribute to the total mass.

In order to make contact with other works in the literature, let us express the mass in  
terms of the original Kaluza-Klein parameters $a$ and $b$. 
Using (\ref{definition of varepsilon}), (\ref{tot grav mass}),  (\ref{inertial mass for Kramer's metric}) and (\ref{m, final expression}) we find
\begin{equation}
\label{masses for N = 1}
M_{g} = M_{in} = m  = {\cal{M}}(a + \frac{b}{2}),
\end{equation}
and in terms of the original Gross-Perry parameters $\alpha$ and $\beta$ (with $\alpha = 1/a, \beta = b/a$)
\begin{equation}
\label{G-P masses for N = 1}
M_{g} = M_{in} = m  = \frac{{\cal{M}}(2 + \beta)}{2 \alpha}.
\end{equation}

\subsection{$N = 0$}

This is the most popular interpretation in the literature, where the metric induced on  the subspace $y =$ constant is interpreted as the effective metric in $4D$. Namely, 
\begin{equation}
\label{exterior metric with N = 0}
ds^2_{(N = 0)} = A^a dt^2 - A^{- (a + b)}dr^2 - r^2 A^{(1 - a - b)}[d\theta^2 + \sin^2\theta d\phi^2].
\end{equation}
This metric has been used in the discussion of many observational problems, which include the classical tests of relativity, as well as the geodesic precession of a gyroscope. It predicts a departure from the equivalence principle, viz.,
\begin{equation}
\label{masses for N = 0}
M_{g} = a {\cal{M}}, \;\;\;M_{in} = m = {\cal{M}}(a + b).
\end{equation}
Notice that the inertial mass in symmetric with respect to $a$ and $b$, which responds to the fact that the spatial part of (\ref{exterior metric with N = 0}) is invariant under $a \leftrightarrow b$. In terms of Gross-Perry parameters
 \begin{equation}
\label{G-P masses for N = 0}
M_{g} = \frac{{\cal{M}}}{\alpha}, \;\;\;M_{in} = m = \frac{{\cal{M}}(1 + \beta)}{\alpha}.
\end{equation}
Thus, only for $b = \beta = 0$, which corresponds to the Schwarzschild vacuum exterior, are the masses equal.

The line element (\ref{exterior metric with N = 0}) is incompatible with the Newtonian limit. Besides, at large distances from the origin the effective matter quantities do not satisfy the dominant energy condition.

\section{The Deser-Soldate definition of mass}

In a well-known paper \cite{Deser}, Deser and Soldate propose a definition for the total energy $P^{0}$, which   explicitly  contains a contribution from the extra dimension (equation $2.7$ in that paper). In our notation, it is 
\begin{equation}
\label{DS mass}
M_{DS} = \frac{1}{16 \pi}\int{\left(\frac{\partial h^{(ind)\;j}_{\;\;\;\;\;\;\;k}}{\partial x^{j}} - \frac{\partial h^{(ind)\;j}_{\;\;\;\;\;\;\;j}}{\partial x^{k}} - \;\frac{\partial h^{5}_{5}}{\partial x^k}\right)\;r^2\;n^{k}d\Omega}.
\end{equation}
In this section we show that Deser-Soldate's mass definition is equivalent to the inertial mass (\ref{total mass}) for the conformal factor $N = 1$. The discussion in this section is general. 

\subsection{General formulae}

Let us apply  (\ref{total mass}) to the case where 
\begin{equation}
\label{effective metric DS}
g_{\mu\nu} \equiv g_{\mu\nu}^{(eff)} = \Phi^{N}g_{\mu\nu}^{(ind)},\;\;\;\;\epsilon \Phi^2 = g_{55},
\end{equation}
where $g_{55}$ is the metric coefficient in front of the extra dimension, and $\epsilon = \pm 1$, depending on whether it is spacelike or timelike\footnote{Do not confuse $\varepsilon$ with $\epsilon$}. Therefore, the Minkowski metric in $5D$ is $\eta_{AB} = diag(1, -1, -1, -1, \;\epsilon)$. Now, setting $g_{55} = h_{55} + \eta_{55}$, the asymptotic behavior, as $r \rightarrow \infty$,  of the effective metric   (\ref{effective metric DS}) is given by
\begin{equation}
g_{\mu\nu} \equiv g_{\mu\nu}^{(eff)} \rightarrow h_{\mu\nu}^{(ind)} + \frac{N}{2}\epsilon\; h_{55}\;\eta_{\mu\nu} + \eta_{\mu\nu},
\end{equation} 
where $h_{\mu\nu}^{ind}$ represents the asymptotic form of the induced metric. In other words, in this limit
\begin{equation}
h _{\mu\nu} \equiv h^{(eff)}_{\mu\nu} \rightarrow h_{\mu\nu}^{(ind)} + \frac{N}{2}\epsilon\; h_{55}\;\eta_{\mu\nu}.
\end{equation}
Consequently,

\begin{equation}
\left(\frac{\partial h^{(eff)\;j}_{\;\;\;\;\;\;k}}{\partial x^j} - \frac{\partial h^{(eff)\;j}_{\;\;\;\;\;\;j}}{\partial x^k}\right) \rightarrow  \left(\frac{\partial h^{(ind)\;j}_{\;\;\;\;\;\;k}}{\partial x^j} - \frac{\partial h^{(ind)\;j}_{\;\;\;\;\;\;j}}{\partial x^k} - N\;\frac{\partial h^{5}_{5}}{\partial x^k}\right).
\end{equation}
Thus, the total inertial mass becomes
\begin{equation}
\label{total mass with h55}
M_{in} = \frac{1}{16 \pi}\int{\left(\frac{\partial h^{(ind)\;j}_{\;\;\;\;\;\;\;k}}{\partial x^{j}} - \frac{\partial h^{(ind)\;j}_{\;\;\;\;\;\;\;j}}{\partial x^{k}} - N\;\frac{\partial h^{5}_{5}}{\partial x^k}\right)\;r^2\;n^{k}d\Omega},
\end{equation}
which depends explicitly on $h_{5}^{5}$ and $N$. First, note  that this result holds regardless of the signature of the extra dimension. Secondly, for $N = 1$, this equation  becomes identical to (\ref{DS mass}), which is the flux integral expression proposed by Deser and Soldate \cite{Deser} . 

\medskip

The above shows that the Deser-Soldate definition of inertial mass should be interpreted as the total inertial mass of classical gravitational systems whose spacetime is conformal to the induced metric, with $\Phi = \sqrt{\epsilon g_{55}}$, i.e. $N = 1$,  as a conformal factor. Indeed, Deser and Soldate indicate that $h_{55}$ can be removed from (\ref{total mass with h55}) through a conformal rescaling of the spacetime (induced) metric, to yield a form like (\ref{total mass}) without the last term. However, they do not elaborate much about this.

\subsection{Example}

Deser and Soldate  applied their definition of total energy to the family of spherically symmetric exterior solutions (\ref{Davidson and Owen solution}) in the form given by Gross and Perry,
\[ 
dS^2 = \left(\frac{1 - {\cal{M}}/2\rho}{1 + {\cal{M}}/2\rho}\right)^{2/\alpha}dt^2 - 
\left(\frac{1 - {\cal{M}}/2\rho}{1 + {\cal{M}}/2\rho}\right)^{2(\alpha - \beta - 1)/\alpha}\left( 1 + \frac{{\cal{M}}}{2\rho}\right)^4[d\rho^2 + \rho^2 d\Omega^2]
\pm  \left(\frac{1 - {\cal{M}}/2\rho}{1 + {\cal{M}}/2\rho}\right)^{2\beta/\alpha}dy^2,
\] 
 and obtained 
\begin{equation}
\label{?corrected version}
M_{DS} = \frac{{\cal{M}}(2 + \beta)}{2\alpha},
\end{equation}
which is identical to (\ref{G-P masses for N = 1}), instead of $M_{in} = {\cal{M}}(1 + \beta)/\alpha$ corresponding to $N = 0$. If we apply the general expression (\ref{total mass with h55}) to Gross-Perry solution we get $M_{DS} = ({\cal{M}}/\alpha)[1 + \beta(1 - N/2)]$, which  becomes identical to (\ref{inertial mass for Kramer's metric}), after changing $\alpha = 1/a, \beta = b/a$. Thus, for $N = 1$
\begin{equation}
M_{g} = M_{in} = m = M_{DS} = \varepsilon {\cal{M}}.
\end{equation}

\medskip

In summary, (\ref{total mass with h55}) gives the inertial mass of a gravitational system whose effective spacetime is conformal to the induced metric, $g_{\mu\nu}^{(eff)} = \Phi^{N}g_{\mu\nu}^{(ind)}$ with $\Phi = \sqrt{\epsilon g_{55}}$. Regarding our particular example, it should be noted that frequently in the literature, (\ref{?corrected version}) is interpreted as the expression for the inertial mass of the effective spacetime  with $N = 0$ (instead of $N = 1$). Our calculations show that such interpretation is inappropriate. 

\section{Summary and concluding remarks}

The $(4 + 1)$ dimensional reduction of the field equations in $5D$ leads to a set of equations in $4D$ which is not closed: the effective equations for gravity in $4D$ are weaker than in ordinary general relativity \cite {Viser}. This reflects the fact that there are many ways of producing, or embedding, a $4D$ spacetime in a given higher-dimensional manifold, while satisfying the field equations \cite{JPdeLgr-qc/0512067}. As a consequence,  the  effective
picture in four dimensions allows the existence of different possible non-Schwarzschild scenarios for the description of the spacetime outside of a spherical star \cite{JPdeLgr-qc/0701129}, \cite{JPdeLarXiv:0711.0998}. 

In this  paper we have examined the equivalence principle mainly in the sense where it refers to the {\it equality of definitions for mass} which are based on a metric, in the tradition of Deser and Soldate \cite{Deser}, Davisson and Owen \cite{Davidson Owen} and Gross and Perry \cite{Gross Perry}. The main question under investigation has been whether the violation of this principle   is a necessary consequence of the existence of extra dimensions. 

In this context, we have shown that the line element 
\begin{equation}
\label{exterior metric with N = 1. Conclusion}
ds^2_{(N = 1)} = \left(1 - \frac{2{\cal{M}}}{r}\right)^\varepsilon dt^2 - \left(1 - \frac{2{\cal{M}}}{r}\right)^{- \varepsilon}dr^2 - \left(1 - \frac{2{\cal{M}}}{r}\right)^{(1 - \varepsilon)}r^2[d\theta^2 + \sin^2\theta d\phi^2],
\end{equation}
is consistent with the weak equivalence principle (\ref{equal masses}), the Newtonian limit,  and the dominant energy condition for any value of the  free parameter $\varepsilon$. In fact, all definitions of mass lead to the same result, namely, $M_{g} = M_{in}  = m = M_{DS} = \varepsilon {\cal{M}}$. We note that the range of values of parameter $\varepsilon$ can be restricted to $0 \leq \varepsilon \leq 1$ if one assumes that the external effective matter density (\ref{exterior matter density}) is positive. For $\varepsilon = 1$ we recover the Schwarzschild solution. We stress the fact that the masses are equal \underline{even in the non-Schwarzschild case} where $\varepsilon \neq 1$. We have not discussed here any particular $5D$ embedding  for this metric but the existence of such an  embedding  is guaranteed by Campbell-Maagard's embedding theorems \cite{SanjeevWesson}, \cite{Indefenceof}. 

Although observations suggest that $\varepsilon$ must be very close to $1$, as a matter of principle let us note that for $\varepsilon = 0$ we get a spacetime whose total inertial and gravitational mass is zero, but which is not flat in $4D$. Indeed, the Riemann tensor, for $\varepsilon = 0$, has the following nonzero components
\begin{equation}
R_{1212} =  \frac{R_{1313}}{\sin^2\theta} = - \frac{{\cal{M}}^2}{r^2 A}, \;\;\;R_{2323} =  {\cal{M}}^2\sin^2\theta. 
\end{equation}
Coming back to our problem, let us notice  that the requirement that in the weak-field limit we recover the usual Newtonian physics, by virtue of  (\ref{compatibility with Newtonian physics}), is equivalent to demand the equality (\ref{equal masses}) among all the masses, as well as the fulfillment of the dominant energy condition (\ref{dom. energy condition}). The opposite is also true, the dominant energy condition demands the equality of masses and  guarantees the validity of the Newtonian limit.  

What this mean is that, at least within the context of Kaluza-Klein theory, the existence of an extra dimension does not necessarily imply a major departure from some basic  principles of general relativity. In fact, the line element  (\ref{exterior metric with N = 1. Conclusion})  is  compatible with  (i) Newtonian physics, in the weak-field limit; (ii) the general-relativistic Schwarzschild limit for $\varepsilon = 1$; (iii) the dominant energy condition, and (iv) the equivalence principle.

The information about the extra dimension is consolidated in the non-local stresses induced in $4D$ from the Weyl curvature tensor in $5D$. Thus, the exterior of a star is not absolutely empty as in the Schwarzschild vacuum exterior, but is surrounded by a cloud of matter (\ref{exterior matter density}) that decreases as $1/r^4$. The relevant characteristic of this cloud is that it has no effect on gravitational interactions. In fact, it satisfies ${(T^0_0 - T^1_1 - T^2_2 - T^3_3)} = 0$, which corresponds to $M_{g} = 0$. In other words, the region outside the surface of a star, described by (\ref{exterior metric with N = 1. Conclusion}),  does not contribute to the total mass, similarly to what happens in ordinary general relativity, where outside the surface $T^0_0 = T^1_1 = T^2_2 = T^3_3 = 0$.  

Thus, the exterior metric (\ref{exterior metric with N = 1. Conclusion}) is 
Schwarzschild-like in many aspects. However, this does not mean that the putative extra dimension has no effects in $4D$. 
\begin{enumerate}
 \item The meaning of the equivalence principle in 5D theories of gravity has been discussed by Wesson \cite{Wesson book}(pp. 85-88). He  pointed out that in more than $4$-dimensions, the definition of the principle becomes semantological, and that the geodesics followed (or not) by test particles may be used to rule out some versions of Kaluza-Klein gravity \cite{Cho}, \cite{Xu}.  
In experiments involving the motion of test particles, in principle, one should be able to measure some anomalous acceleration. From (\ref{g for moving particles}), with $|V| \ll 1$ and $N = 1$, we get

\begin{equation}
\label{anomalous acceleration}
g = - \frac{{\cal{M}}}{r^2 \sqrt{A}}\left\{1 + \frac{\ln{A} + 2}{2}(\varepsilon - 1) + 
 \frac{\ln{A}(\ln{A} + 4)}{8}(\varepsilon - 1)^2 + O\left((\varepsilon - 1)^3)\right)\right\},
\end{equation}
which an observer in $4D$ \underline{could} interpret as a violation of the equivalence principle. However, this is not a violation of  (\ref{equal masses}) but a departure from the usual $(4D)$ law of geodesic motion, which can be  interpreted as some sort of non-gravitational force\footnote{Like the so-called ``fifth" force} (In this sense ``the problem is largely one of semantics" \cite{Wesson book}).

\item In the context of the stellar structure, the deviation from the Schwarzschild vacuum exterior affects the parameters of a neutron star. The general relativity upper  limit $M/R < 4/9$ is significantly {\it increased} as we go away from the Schwarzschild vacuum exterior. In principle, the compactness limit of a star can be larger than $1/2$, without being a black hole. Again, more work is needed in this area. 

\end{enumerate}

Also, we have clarified the meaning of the Deser-Soldate definition of total energy, which is usually interpreted as the total mass (or energy) of a spacetime whose line element is the induced metric, i.e., $g_{\mu\nu} = g_{\mu\nu}^{(ind)}$.  We have shown that it corresponds to a spacetime with metric $g_{\mu\nu} = \sqrt{\varepsilon g_{55}} \;g_{\mu\nu}^{(ind)}$.

 We would like to finish this paper with the following remarks. Theories in $5D$ allow the existence of different, non-Schwarzschild, scenarios for the description of the spacetime outside of a spherical star, contrary to four-dimensional general relativity, where we have only one possibility which is the Schwarzschild exterior metric. Then, an observer in $4D$, who is not directly aware of the existence of an extra dimension, will interpret these non-Schwarzschild exteriors as if they were governed by an effective energy-momentum tensor.   

For example, the vacuum solutions discussed in references \cite{Cristiano}-\cite{Bronnikov}, have non-vanishing effective energy-momentum tensors, which are (as usually) defined through the conventional Einstein equations. In \cite{Bronnikov}, among other things,  the authors present the explicit form of the effective matter quantities for various non-Schwarzschild vacuum exteriors. These are not zero, in particular $\rho_{out} \neq 0$, but are ``quite exotic" from the viewpoint of the energy conditions. Certainly, they do not represent any kind of ``regular" (or baryonic) matter that we know of; their nature is purely geometrical. 

In the case of the Kaluza-Klein models under consideration we have seen in section $4$ that the dominant energy condition is  {\it not} an additional assumption but a consequence of the fulfillment of the equivalence principle. The effective matter outside of the star is gravitationally innocuous, in the sense that the corresponding gravitational mass is zero (equations (\ref{standard gravitational mass}) and (\ref{exotic exterior})) as if there were no matter at all.  However, as discussed above in (\ref{anomalous acceleration}) it should affect the motion of test particles as well as the stellar structure \cite{JPdeLgr-qc/0701129}.

\medskip

\paragraph{Acknowledgment:} I would like to thank Paul Wesson for his comments on the meaning of the equivalence principle in more than $4$-dimensions.

\renewcommand{\theequation}{A-\arabic{equation}}
  \setcounter{equation}{0}  
  \section*{Appendix A: Boundary conditions}  
In general relativity, a covariant presentation of the matching conditions, across a separating hypersurface,  requires the continuity of the first and second fundamental forms. However, for  every particular line element  these conditions might look quite differently. Here we present them for the  metrics under consideration.

The interior of a spherical star will be described by the static line element
\begin{equation}
\label{static interior, BC}
ds^2 = e^{\omega(R)}dt^2 - e^{\sigma(R)}dR^2 - R^2 [d\theta^2 + \sin\theta d\phi^2],
\end{equation}
while the spacetime outside of a star is assumed to pertain to the family given by (\ref{eff. spacetime for Kramer's metric}), which has the form 
\begin{equation}
\label{general exterior metric, BC}
ds^2 = e^{\nu(r)}dt^2 - e^{\lambda(r)}dr^2 - r^2 e^{\mu(r)}[d\theta^2 + \sin^2\theta d\phi^2].
\end{equation}
The exterior boundary of the star is a three-dimensional surface $\Sigma$ defined as $R = R_{b}$, and $r = r_{b}$ from inside and outside, respectively.

Standard matching conditions require continuity of the metric at $\Sigma$. Namely,

\begin{equation}
\label{continuity of the first fundamental form}
e^{\omega(R_{b})} = e^{\nu(r_{b})},\;\;\;R_{b} = r_{b}e^{\mu(r_{b})/2}.
\end{equation}
If $n_{\mu} = \delta_{\mu}^{1}\sqrt{- g_{11}}$ represents the unit vector orthogonal to the boundary surface,  then the second fundamental form  (say $dK^2$) calculated from inside and evaluated at 
$\Sigma_{|R_{b}} $ gives 
\begin{equation}
\label{seconf fund. form from inside}
dK^2_{|\Sigma_{R_{b}} } = - e^{- \sigma(R_b)/2}\left\{\frac{1}{2}\omega_{R}(R_{b})e^{\omega(R_{b})}\;dt^2 - R_{b}[d\theta^2 + \sin^2\theta \; d\phi^2] \right\}.
\end{equation} 
From outside yields
\begin{equation}
\label{seconf fund. form from outside}
dK^2_{|\Sigma_{r_{b}} } = - e^{- \lambda(r_b)/2}\left\{\frac{1}{2}\nu_{r}(r_{b}) e^{\nu(r_{b})}\;dt^2 - \left(\frac{1}{r_{b}} + \frac{1}{2} \mu_{r_{b}}(r_{b})\right)r_{b}^2e^{\mu(r_{b})}[d\theta^2 + \sin^2\theta \; d\phi^2] \right\}
\end{equation} 
Now, the continuity of the second fundamental form requires
\begin{equation}
\label{cont. of sec. fund. form 1}
e^{-\sigma(R_{b})/2}\omega_{R}(R_{b}) = e^{- \lambda(r_{b})/2}\nu_{r}(r_{b}),
\end{equation}
and 
\begin{equation}
\label{cont. of sec. fund. form 2}
e^{- \sigma(R_{b})/2} = e^{- \lambda(r_{b})/2}\left(\frac{1}{r_{b}} + \frac{\mu_{r}(r_{b})}{2}\right)r_{b} e^{\mu(r_{b})/2}.
\end{equation}

Inside the fluid sphere the gravitational mass (within a sphere of radius $R$) and the radial ``pressure" $p_{r} = = - T_{1}^{1} $ are given by 

\begin{equation}
\label{grav. mass for the Schw. interior}
M_{g}(R) = \frac{1}{2}R^2e^{(\omega - \sigma)/2}\omega_{R},
\end{equation}
\begin{equation}
\label{internal radial pressure} 
8 \pi p_{r} = e^{- \sigma}\left(\frac{\omega_{R}}{R} + \frac{1}{R^2}\right) - \frac{1}{R^2}.
\end{equation}
Meanwhile, outside of the sphere we have
\begin{equation}
M_{g}(r) = \frac{1}{2}r^2e^{(\mu + \nu - \lambda)/2}\nu_{r},
\end{equation}
and
\begin{equation}
8 \pi p_{r} = e^{- \lambda}\left(\frac{\nu_{r} + \mu_{r}}{r} + \frac{\nu_{r} \mu_{r}}{2} + \frac{\mu_{r}^2}{4} + \frac{1}{r^2}\right) - \frac{e^{- \mu}}{r^2}.
 \end{equation}
It is clear that the continuity of the first and second fundamental forms across the boundary surface guarantees the continuity of the gravitational mass and radial pressure.

\renewcommand{\theequation}{B-\arabic{equation}}
  \setcounter{equation}{0}  
  \section*{Appendix B: Definition of total energy}  

When applying the formalisms of general relativity one has to take special care of conventions and definitions. Some authors work with signature ($+, -, -, -$), and others with $(-, +, +, +)$. Also, there are different definitions for  the Riemann-Christoffel curvature tensor. As a consequence the Einstein field equations look different. For example, $G_{\mu\nu} = - 8\pi G T_{\mu\nu}$  in \cite{Weinberg} and $G_{\mu\nu} = 8\pi G T_{\mu\nu}$ in \cite{Landau}. Besides, the quantity $Q^{\alpha\beta\gamma}$ (see bellow) is defined  with different symmetry properties. The object of this appendix is to make sure we are using the ``correct" equations. Therefore, we follow Weinberg's neat presentation \cite{Weinberg}, but use the conventions of Landau and Lifshitz \cite{Landau}.

Let us start by decomposing the spacetime metric into its asymptotic value $\eta_{\mu\nu}$, which is the Minkowski metric (\ref{Minkowski metric}),  and a deviation $h_{\mu\nu}$, viz.,  
\begin{equation}
\label{definition of h}
g_{\mu\nu} = h_{\mu\nu} + \eta_{\mu\nu}.
\end{equation}
The part of the Ricci tensor linear in $h_{\mu\nu}$ is 
\begin{equation}
\label{R1}
R_{\;\;\;\;\mu\nu }^{(1)} = \frac{1}{2}\left(\frac{\partial^{2}h_{\mu}^{\alpha}}{\partial x^{\nu}\partial x^{\alpha}} + \frac{\partial^{2}h_{\nu}^{\alpha}}{\partial x^{\mu}\partial x^{\alpha}} -  \frac{\partial^{2}h^{\alpha}_{\alpha}}{\partial x^{\mu}\partial x^{\nu}} - \eta^{\alpha\beta}\frac{\partial^{2}h_{\mu\nu}}{\partial x^{\alpha}\partial x^{\beta}}\right),
\end{equation} 
where $h_{\beta}^{\alpha} = \eta^{\alpha \rho} h_{\rho \beta}$ and $h^{\alpha}_{\alpha} = \eta^{\alpha\beta}h_{\alpha\beta}$. Now,
\begin{equation}
\label{Einstein, first order in h}
R_{\;\;\;\;\mu\nu}^{(1)} - \frac{1}{2}\;\eta_{\mu\nu}\;R^{(1)} = 8\pi\;\left(T_{\mu \nu} + t_{\mu\nu}\right),
\end{equation}
with $R^{(1)} = \eta^{\mu\nu}R_{\;\;\;\;\mu\nu}^{(1)}$ and 
\begin{equation}
t_{\mu\nu} = \frac{1}{8\pi}\left[\left(R_{\;\;\;\;\mu\nu}^{(1)} - \frac{1}{2}\;\eta_{\mu\nu}\;R^{(1)}\right) - \left(R_{\mu\nu} - \frac{1}{2}\;g_{\mu\nu}\;R\right)\right],
\end{equation}
where we have used the field equations
\begin{equation}
R_{\mu\nu} - \frac{1}{2}\;g_{\mu\nu}\;R = 8\pi T_{\mu \nu}.
\end{equation}
Let us denote
\begin{equation}
G^{(1)\mu}_{\;\;\;\;\;\nu} \equiv R_{\;\;\;\;\;\nu}^{(1)\mu} - \frac{1}{2}\;\delta^{\mu}_{\nu}\;R^{(1)}. 
\end{equation}
Thus, using (\ref{R1})
\begin{equation}
\label{general G1}
G^{(1)\mu\nu}  = 
\frac{1}{2}\left(\eta^{\mu\alpha}\frac{\partial^{2}h^{\nu\beta}}{\partial x^{\alpha}\partial x^{\beta}} + \;\eta^{\nu\beta}\;\frac{\partial^{2}h^{\mu \alpha}}{\partial x^{\alpha}\partial x^{\beta}} + \eta^{\mu\nu}\;\eta^{\alpha\beta}\frac{\partial^{2}h_{\lambda}^{\lambda}}{\partial x^{\alpha}\partial x^{\beta}} - \eta^{\mu\alpha}\eta^{\nu\beta}\frac{\partial^{2}h_{\lambda}^{\lambda}}{\partial x^{\alpha}\partial x^{\beta}} - \eta^{\alpha\beta}\frac{\partial^{2}h^{\mu\nu}}{\partial x^{\alpha}\partial x^{\beta}} - \eta^{\mu\nu}\frac{\partial^{2}h^{\alpha\beta}}{\partial x^{\alpha}\partial x^{\beta}}\right).
\end{equation}
Notice that 
\begin{equation}
\label{Conservation equation up to h}
\frac{\partial }{\partial x^{\nu}} \;G^{(1)\mu\nu} = \frac{\partial  }{\partial x^{\mu}}\; G^{(1)\mu\nu} = 0.
\end{equation}
Therefore, the r.h.s. of (\ref{Einstein, first order in h})
\begin{equation}
{\cal{T}}^{\mu\nu} = \eta^{\mu\lambda}\eta^{\nu\rho}\;[T_{\lambda\rho} + t_{\lambda\rho}]
\end{equation}
is a locally conserved quantity, i.e.,
\begin{equation}
\label{locally conserved quantity}
\frac{\partial}{\partial x^{\mu} } \;{\cal{T}}^{\mu\nu}= \frac{\partial}{\partial x^{\nu}}{\cal{T}}^{\mu\nu} = 0.
\end{equation}
There are a number of reasons \cite{Weinberg}  which suggest that  ${\cal{T}}^{\mu\nu}$ should be  interpreted as the {\it total} energy-momentum  pseudo-tensor of matter and gravitation, and $t_{\mu\nu}$ as the energy-momentum  pseudo-tensor of the gravitational field itself \cite{Landau}. 

In addition the derivatives $(\partial/\partial x^{\alpha})$ and $(\partial/\partial x^{\beta})$ appear in all terms of (\ref{general G1}). Then, by virtue of the constancy of the metric coefficients  $\eta_{\mu\nu}$, one can factorize one, or both, of them\footnote{As is done in Landau and Lifshitz \cite{Landau}, where they introduce the quantity $\lambda^{\mu\nu\alpha\beta}$.}. The theoretical working looks a little different, but the final result is the same.

\paragraph{Factorizing $\partial/\partial x^{\alpha}$:} If we choose to factorize $(\partial/\partial x^{\alpha})$, then we obtain
\begin{equation}
\label{G1 for QLT}
G^{(1)\mu\nu}  =  8\pi {\cal{T}^{\mu\nu}} = \frac{\partial }{\partial x^{\alpha}}\;Q^{\mu\nu\alpha}_{LT}\;,
\end{equation}
with

\begin{equation}
\label{definition of QLT}
Q^{\mu\nu\alpha}_{LT} = \frac{1}{2}\left(\eta^{\mu\alpha}\frac{\partial h^{\nu\beta}}{\partial x^{\beta}} + \;\eta^{\nu\beta}\;\frac{\partial h^{\mu \alpha}}{\partial x^{\beta}} + \eta^{\mu\nu}\;\eta^{\alpha\beta}\frac{\partial h_{\lambda}^{\lambda}}{\partial x^{\beta}} - \eta^{\mu\alpha}\eta^{\nu\beta}\frac{\partial h_{\lambda}^{\lambda}}{\partial x^{\beta}} - \eta^{\alpha\beta}\frac{\partial  h^{\mu\nu}}{\partial x^{\beta}} - \eta^{\mu\nu}\frac{\partial h^{\alpha\beta}}{\partial x^{\beta}}\right).
\end{equation}
Here $LT$ means that this quantity is antisymmetric in its last $(L)$ two $(T)$ indices. Namely,
\begin{equation}
Q^{\mu\nu\alpha}_{LT} = - Q^{\mu\alpha\nu}_{LT}.
\end{equation}
Thus, $\partial G^{(1)\mu\nu}/\partial x^{\nu} = 0 = \partial {\cal{T}^{\mu\nu}}/\partial x^{\nu}$ is satisfied automatically.

\paragraph{Factorizing $\partial/\partial x^{\beta}$:} If we choose to factorize $(\partial/\partial x^{\beta})$, then we obtain
\begin{equation}
\label{G1 for QFT}
G^{(1)\mu\nu}  =  8\pi{\cal{T}^{\mu\nu}}  = \frac{\partial }{\partial x^{\beta}}\;Q^{\beta\mu\nu}_{FT},
\end{equation}
with

\begin{equation}
\label{definition of FT}
Q^{\beta\mu\nu}_{FT} = \frac{1}{2}\left(\eta^{\mu\alpha}\frac{\partial h^{\nu\beta}}{\partial x^{\alpha}} + \;\eta^{\nu\beta}\;\frac{\partial h^{\mu \alpha}}{\partial x^{\alpha}} + \eta^{\mu\nu}\;\eta^{\alpha\beta}\frac{\partial h_{\lambda}^{\lambda}}{\partial x^{\alpha}} - \eta^{\mu\alpha}\eta^{\nu\beta}\frac{\partial h_{\lambda}^{\lambda}}{\partial x^{\alpha}} - \eta^{\alpha\beta}\frac{\partial  h^{\mu\nu}}{\partial x^{\alpha}} - \eta^{\mu\nu}\frac{\partial h^{\alpha\beta}}{\partial x^{\alpha}}\right).
\end{equation}
This quantity is antisymmetric in its first (F) two (T)indices,
\begin{equation}
Q^{\beta\mu\nu}_{FT} = - Q^{\mu\beta\nu}_{FT}.
\end{equation}
Thus, $\partial G^{(1)\mu\nu}/\partial x^{\mu} =  0 = \partial {\cal{T}^{\mu\nu}}/\partial x^{\mu}$ is satisfied automatically.

From (\ref{locally conserved quantity}) it follows that quantity
\begin{equation}
P^{\mu} = \int_{V}{{\cal{T}}^{\mu 0}dx^3}
\end{equation}
may be interpreted as the total energy-momentum pseudo-vector of the system including matter, all non-gravitational fields, and gravitation. Using (\ref{G1 for QLT}) or (\ref{G1 for QFT}) this becomes
\begin{equation} 
P^{\lambda} = \frac{1}{8\pi}\int_{V}{\frac{\partial }{\partial x^{\beta}}\;Q^{\beta 0\lambda}_{FT}\;dx^3} = \frac{1}{8\pi}\int_{V}{\frac{\partial }{\partial x^{\alpha}}\;Q^{\lambda 0\alpha}_{LT}\;dx^3}.
\end{equation}
Using Gauss's theorem, integrating over a large sphere of radius $r$ we get
\begin{equation} 
P^{\lambda} = - \frac{1}{8\pi}\int{Q^{j 0\lambda}_{FT}\; r^2\;n_{j}d\Omega} = - \frac{1}{8\pi}\int{Q^{\lambda 0\j}_{LT}\;r^2\;n_{j}d\Omega},
\end{equation}
where $d\Omega = \sin\theta d\theta d\phi$ and $n^{k}$ is the unit vector pointing outward
\begin{equation}
n^{k} = \frac{x^{k}}{r}, \;\;\;\eta_{jk}n^{i}n^{k} = - 1.
\end{equation}
The total inertial mass $M_{in}$ is given by
\begin{equation} 
M_{in} =  P^{0} = - \frac{1}{8\pi}\int{Q^{j 0 0}_{FT}\; r^2\;n_{j}d\Omega} = - \frac{1}{8\pi}\int{Q^{0 0 j}_{LT}\;r^2\;n_{j}d\Omega}.
\end{equation}
After a simple calculation we obtain
\begin{equation}
\label{total mass}
M_{in} = \frac{1}{16 \pi}\int{\left(\frac{\partial h^{j}_{k}}{\partial x^{j}} - \frac{\partial h^{j}_{j}}{\partial x^{k}}\right)\;r^2\;n^{k}d\Omega}.
\end{equation}

\end{document}